\documentclass[runningheads]{llncs}

\usepackage[utf8]{inputenc}
\usepackage{subcaption}
\usepackage{multirow}
\usepackage{tabularx, booktabs}
\newcolumntype{Y}{>{\centering\arraybackslash}X}
\usepackage{amsmath}
\usepackage{url}
\usepackage{xspace}
\usepackage{multirow}
\usepackage{adjustbox}
\usepackage{cite}
\usepackage[font={small}]{caption}
\usepackage{graphicx}
\usepackage{amsfonts}
\DeclareGraphicsExtensions{.pdf}

\usepackage{hyperref}

\hypersetup{
    pdfcreator={},
    pdfproducer={}
}

\newcommand{\etal}{{\it et al}.\xspace}
\newcommand{\ie}{{\it i.e.},\ }
\newcommand{\eg}{{\it e.g.},\ }

\newcolumntype{L}[1]{>{\raggedright\let\newline\\\arraybackslash\hspace{0pt}}m{#1}}
\newcolumntype{C}[1]{>{\centering\let\newline\\\arraybackslash\hspace{0pt}}m{#1}}
\newcolumntype{R}[1]{>{\raggedleft\let\newline\\\arraybackslash\hspace{0pt}}m{#1}}

\newcommand{\algoname}{{${\textrm{TAH}}^{\prime}$}\xspace}
\newcommand{\fuzzname}{{TAH}\xspace}
\newcommand{\B}[1]{\texttt{B\textsubscript{#1}}}

\begin{document}



\date{}

\title{Topology-Aware Hashing for\\Effective Control Flow Graph Similarity
  Analysis}
\titlerunning{Topology-Aware Hashing for Effective CFG Similarity Analysis}
%
\author{Yuping Li\inst{1}, Jiyong Jang\inst{2}, \and Xinming Ou\inst{3}}
\institute{Pinterest \and IBM Research \and University of South Florida}

\maketitle
\begin{abstract}

Control Flow Graph (CFG) similarity analysis is an essential technique for a
variety of security analysis tasks, including malware detection and malware clustering.
Even though various algorithms have been developed, existing CFG similarity
analysis methods still suffer from limited efficiency, accuracy, and usability.
In this paper, we propose a novel fuzzy hashing scheme called
topology-aware hashing (\fuzzname)
for effective and efficient CFG similarity analysis.
Given the CFGs constructed from program binaries, we extract blended
$n$-gram graphical features of the CFGs, encode the graphical features
into numeric vectors (called graph signatures), and then
measure the graph similarity by comparing the graph signatures.
We further employ a fuzzy hashing technique to convert the numeric graph
signatures into smaller fixed-size fuzzy hash signatures for efficient
similarity calculation.
Our comprehensive evaluation demonstrates that \fuzzname is more
effective and efficient compared to existing CFG comparison techniques.
To demonstrate the applicability of \fuzzname to real-world security analysis
tasks,
we develop a binary similarity analysis tool based on \fuzzname,
and show that it outperforms existing similarity analysis tools
while conducting malware clustering.


\end{abstract}




\section{Introduction}

Control flow graph (CFG) similarity analysis has played an essential role in malware analysis, \eg
detecting the variants of known malware samples~\cite{hu2009large,dullien2005graph,
  kruegel2005polymorphic, cesare2011malware, bourquin2013binslayer}, evaluating the
relationship between different malware families, studying the evolution of
different malware families, and triaging large-scale newly
collected malicious samples to prioritize the new threats.
Research also demonstrated that it is the most fundamental
component for effective binary bug search~\cite{eschweiler2016discovre}, which tried to
create signatures from known vulnerable version CFGs and
identify the vulnerable version of binaries.
Therefore, effective and efficient CFG similarity analysis is much desired
for operational security analysis in practice.

Despite the numerous efforts towards effective CFG similarity comparison,
we found it is still challenging to apply existing CFG comparison approaches
for real-world analysis (\eg structural based binary similarity analysis).
Graph matching is known to be computationally expensive.
Even though several approximate graph isomorphism
algorithms~\cite{mcgregor1982backtrack,
kruegel2005polymorphic,hu2009large,vujovsevic2013software} have been
developed and applied to CFG comparison over the past several decades,
it is still a time-consuming procedure to compare a large number of CFGs at the same time.
For instance, comparing binary {\tt A} with {\it m} functions and binary
{\tt B} with {\it n} functions would result in $m*n$ pairwise CFG comparisons.
In addition, we notice that the majority of existing CFG similarity
comparison algorithms work with ``raw'' CFG structures,
and rely on inefficient CFG representations for comparison.
Last but not least, the evaluation of existing CFG comparison algorithms
mainly focus on recognizing the similarity of CFGs.
However, performing well with regard to recognizing CFG similarities does
not guarantee also doing good in identifying CFG differences.
And the later capability is equally important especially if the information
of how ``similar'' of two CFGs are also used in practical applications,
\eg whether the algorithm will generate comparatively low similarity
scores if the input CFGs are significantly different.


In this paper, we hypothesize that the original CFG representation is not required to measure
the CFG similarity if a CFG can be effectively
encoded with certain representative graph features, which could result
in a universal and compact graph format and make the overall comparison
more efficient at the same time.
Therefore, based on the insight
that the $n$-gram concept is applicable to represent CFGs to assess
graph similarity, we design a blended $n$-gram graphical feature based
CFG comparison method, called topology-aware hashing (\fuzzname).
The $n$-gram concept has been extensively applied for measuring document
similarity where
contiguous sequences of {\it n} items are extracted from an input stream.
We apply it to CFGs in a similar manner, except working with multiple input paths.
Extracting $n$-gram graphical features from CFG structures enables us to
effectively encode arbitrary CFGs to the same format.
To facilitate the comparison between graph
signatures, we further employ a fuzzy hashing technique to convert the numeric graph
signatures into smaller fixed-size fuzzy hash signatures. In this way, we
achieve high accuracy through the $n$-gram graphical feature
representation, and high efficiency through the compact fuzzy hash comparison.
Compared to the state-of-the-art CFG comparison algorithms,
our approach achieves the highest accuracy for hierarchical clustering
and takes the least amount of time to complete all pairwise comparisons.
To demonstrate the effectiveness of the structural comparison approach,
we further implement a binary similarity analysis tool based on \fuzzname.
When compared with the state-of-the-art binary similarity tools,
it achieves the highest accuracy at the F-score of 0.929 for singe-linkage
malware clustering tasks.


In summary, we have the following major contributions:
\begin{itemize}

\item We propose a blended $n$-gram graphical feature based CFG comparison
  method called \fuzzname. It extracts the $n$-gram graphical features from
  the topology of CFGs, and measures the similarity of CFGs by comparing the graphical
  features encoded in fuzzy hash signatures.

\item We design a clustering analysis based evaluation framework to comprehensively
  assess various CFG comparison techniques, and show that
  \fuzzname is more stable, faster, and generates more accurate results
  compared to state-of-the-art CFG comparison techniques.

\item We design and implement a \fuzzname-based binary similarity analysis tool,
  and demonstrate that it effectively performs malware
  clustering tasks with 2865 carefully labeled malware samples in an efficient manner.


\end{itemize}



\section{Related Work}
\label{sec:related}

\subsection{CFG Similarity Analysis}

Control flow graph (CFG) similarity analysis is the core technical component of many
existing security analysis systems,
and various techniques have been proposed for approximate CFG similarity
computation.

\begin{enumerate}

\item {\bf Min-cost bipartite graph matching:} Hu~\etal~\cite{hu2009large}
  developed an edit distance based graph isomorphism algorithm
by building a cost matrix that represents the costs of mapping the
nodes in two graphs, and using the Hungarian algorithm~\cite{kuhn1955hungarian}
to find an optimal mapping between the nodes such that the total cost (\ie edit
distance) is minimized.
Vujošević~\etal~\cite{vujovsevic2013software} iteratively built a similarity
matrix
between the nodes of two CFGs based on the similarity of their neighbors, and
adopted the Hungarian algorithm to find the matching between the nodes in two
graphs such that the resulting similarity score is the highest.


\item {\bf Maximal common subgraph matching:}
  McGregor~\cite{mcgregor1982backtrack} designed a backtrack search
algorithm to find the maximal common subgraph of two graphs.
This idea has been used to design efficient CFG comparison algorithms, and
adopted for binary semantic difference
analysis~\cite{gao2008binhunt} and  binary code
search~\cite{eschweiler2016discovre} scenarios.
Given the maximal common subgraph output, a graph similarity score was
calculated
as the maximal number of common subgraph nodes divided by the number of
available nodes between two graphs.

\item {\bf $k$-subgraph matching:} Kruegel \etal~\cite{kruegel2005polymorphic}
  designed an algorithm based on $k$-subgraph mining.
They generated a spanning tree for each node in the graph such that the out-degree
of every node was less than or equal to 2, then recursively generated
$k$-subgraphs from the spanning trees by considering all possible allocations of
$k-1$ nodes under the root node.
Each $k$-subgraph was then canonicalized and converted into a fingerprint
by concatenating the rows of its adjacency matrix.


\item {\bf Simulation-based graph similarity:}
  Sokolsky~\etal~\cite{sokolsky2006simulation}
modeled the control flow graphs using Labeled Transition Systems.
Given two CFGs, they recursively matched the most similar outgoing
nodes starting from the entry nodes, and summed up the similarity of the
matched nodes and edges. The overall similarity of two CFGs was then defined by
a recursive formula.

\item {\bf Graph embedding:}
Genius~\cite{feng2016scalable} was designed to learn high-level feature
representations from an attributed CFG (ACFG) and encode the graphs into numerical vectors
using a codebook-based graph matching approach.
It used 6 block-level attributes (\eg string constants and the number of
instructions) and 2 inter-block level attributes (\eg the number of offspring
and betweenness).
Using the same features, Gemini~\cite{xu2017neural} proposed a neural
network-based approach to compute the graph embedding for an ACFG, and achieved
better accuracy and efficiency.
CFG similarity is then measured by comparing the embedded graph representation.


\end{enumerate}

Our \fuzzname algorithm
belongs to the graph embedding category, which is also
known as topological
descriptors~\cite{galvez1994charge,hert2004comparison} in other domains.
We notice that lots of recent graph embedding techniques~\cite{goyal2018graph}
were mainly designed to represent the individual graph nodes~\cite{ahmed2013distributed,ou2016asymmetric} in vector spaces.
They were often applied for network (\ie undirected graph) structure
analysis~\cite{wang2016structural} and require additional training
processes~\cite{wang2016structural,cao2016deep,kipf2016semi}.
\fuzzname is different from Genius
and Gemini in that \fuzzname is basic block content-agnostic, and
its graph embedding is always deterministic and requires no separate training
process.
%
%
Graph kernels are widely adopted
for comparing graph similarities.
However, we note that the majority of the graph kernels
are designed and used for analyzing undirected graphs or
networks\cite{gartner2003survey,
  vishwanathan2010graph,shervashidze2011weisfeiler},
and require label~\cite{shervashidze2011weisfeiler} and
weight~\cite{kondor2016multiscale} information.
Therefore, they are not directly applicable to analyzing CFGs, which
are directed, unlabeled, and unweighted graphs.
Nevertheless, we notice that our $n$-gram concept resonates some of the
structural properties used in graph kernel algorithms,
such as graphlets~\cite{
shervashidze2011weisfeiler} (e.g., the subgraphs
with {\it k} nodes where $k \in {3,4,5}$).

\subsection{Binary Similarity Analysis}
We discuss the previous binary similarity analysis methods that are
most relevant to our approach.

BitShred~\cite{jang2011bitshred} was a system designed for large scale malware
similarity analysis and clustering, and it extracted $n$-gram features from the
machine code sequences of the executable sections and applied
feature hashing to encode the features into a bit-vector.
nextGen-hash~\cite{li2015experimental} was a concretized fuzzy hashing approach
based on the core ideas developed in BitShred, and achieved more accurate
results than other fuzzy hash algorithms; however, its
significant fingerprint size made it hard to use in practice.
Myles and Collberg~\cite{myles2005k} proposed to use opcode-level $n$-grams as
software birthmarks and applied it to prove the copyright of software.
Opcode level $n$-gram representation of a binary has also been explored to
detect similar malicious code patterns~\cite{moskovitch2008unknown,
  shabtai2009detection, hu2013duet, canfora2015effectiveness}.
Alazab~\etal~\cite{alazab2010towards} proposed to detect malware using $n$-gram
features from API call sequences.
SSdeep~\cite{kornblum2006identifying} was a representative fuzzy hashing
algorithm that was used to detect homologous files using context triggered piecewise hashes.


CFG based analysis was also used for binary code comparison.
BinDiff~\cite{dullien2005graph} was a binary comparison tool that assisted
vulnerability
researchers and engineers to quickly find the differences and similarities
using function and basic block level attributes.
BinSlayer~\cite{bourquin2013binslayer} modeled a binary diffing problem
as a bipartite graph matching problem. It assigned a distance
metric between the basic block in one function and the basic block in another
function that minimized the total distance, and found that graph isomorphism
based algorithms were less accurate when the changes between two binaries were large.
BinHunt~\cite{gao2008binhunt} and iBinHunt~\cite{ming2012ibinhunt} relied on
symbolic execution and a theorem prover to check semantic differences between
basic blocks. Although they might yield better accuracy, it was hard to use
in practice due to an expensive operation cost.
Kruegel~\etal~\cite{kruegel2005polymorphic} used the previously mentioned
$k$-subgraph matching algorithm to detect polymorphic worms, which was also
based on CFG structural analysis.
Cesare and Xiang~\cite{cesare2011malware} also extracted fixed-size
$k$-subgraph features and $n$-gram features from the string representation of a CFG
for malware variant detection.


To some extent, our $n$-gram graphical features are close to
$k$-subgraphs, but they are different concepts.
Unlike existing work that tried to find an optimal matching
between functions in two binaries, \fuzzname compares
the CFGs of the entire binary using the overall $n$-gram graphical features.
Furthermore, $n$-gram features from a CFG string~\cite{cesare2011malware} were
still derived in a traditional $n$-gram usage manner
while our proposed $n$-gram graphical features are directly extracted from CFG
structures.


\section{Approach Overview}
\label{sec:overview}

\begin{figure}[ht]
  \centering
  \includegraphics[width=0.9\linewidth]{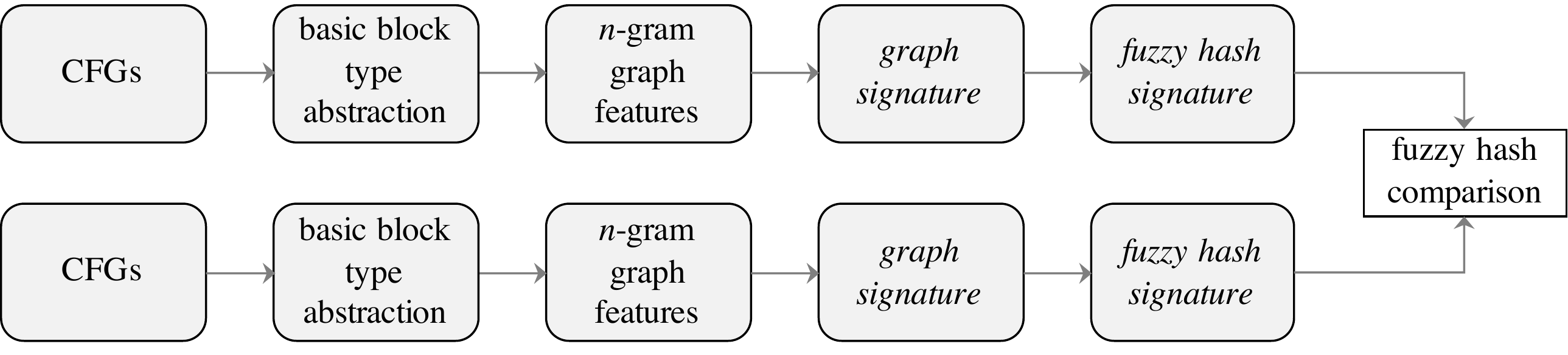}
  \centering
  \caption{The workflow of \fuzzname}
  \label{fig:overview}
\end{figure}

We illustrate the workflow of \fuzzname in Figure~\ref{fig:overview}.
Given two sets of CFGs, we extract the blended $n$-gram graphical features from the
input CFGs and encode them as numeric vectors, called graph signatures.
To make it more efficient to use and compare,
we subsequently convert the graph signatures into fixed-size
bit-vectors, called fuzzy hash outputs.
Finally, we compare the corresponding fuzzy hash outputs
to calculate the similarity of input CFGs.

\subsection{Basic Block Type Abstraction}
\label{subsec:type}


In order to extract representative graphical features, we abstract the basic
blocks of CFGs using categorization.
The main objective of the abstraction is to categorize the nodes of CFGs into
different types, which are then used to denote the ``content'' of the nodes
as used in traditional $n$-gram application scenarios.


We explore a simple yet effective abstraction of basic block types
which captures the topology of a CFG, and demonstrate
that such a simple type abstraction approach produces reliable results.
In particular, we define the basic block types based on the number of parents
(\ie node in-degree) and the number of children (\ie node out-degree).
To study the CFGs of real-world applications including goodware and malware,
we experimentally
analyzed a total of 93,470 binaries that were obtained from
newly installed Android and Windows operation systems, and
malware sharing websites like VirusShare~\cite{virusshare}.


\begin{figure}
  \centering
    \begin{minipage}{0.33\linewidth}
        \centering
        \scalebox{0.95}{
\begin{tabular}{|c|c||c|c|c|c|}
\hline
\multicolumn{2}{|c||}{}& \multicolumn{4}{c|}{outdegree} \\  \cline{3-6}
\multicolumn{2}{|c||}{}& 0 & 1 & 2 & $\geq$3 \\
\hline
\hline
\multirow{4}{*}{indegree} &0 & \B{00} & \B{01} & \B{02} & \B{03} \\
&1 & \B{10} & \B{11} & \B{12} & \B{13} \\
&2 & \B{20} & \B{21} & \B{22} & \B{23} \\
&$\geq$3 & \B{30} & \B{31} & \B{32} & \B{33} \\
\hline
      \end{tabular}
       }
    \end{minipage}
    \begin{minipage}{0.33\linewidth}
        \centering
        \includegraphics[height=0.13\textheight]{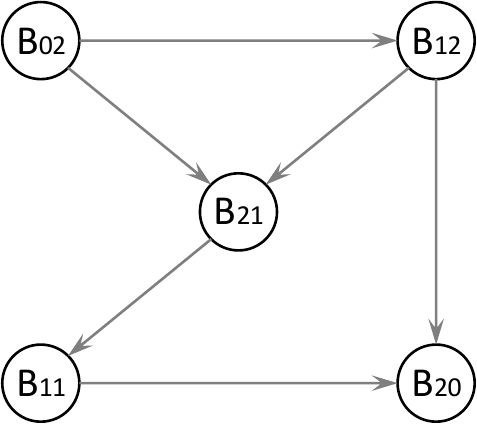}
    \end{minipage}%
    \begin{minipage}{0.33\linewidth}
        \centering
        \scalebox{0.95}{
        \scriptsize
      \begin{tabular}{|c|l|}
        \hline
        $n$ & {$n$-gram feature set} \\
        \hline
        \hline
        1 & {\B{02}, \B{12}, \B{21}, \B{11}, \B{20}} \\
        \hline
        2 & {\B{02}\B{12}, \B{02}\B{21}, \B{12}\B{21}, \B{12}\B{20}} \\ & {\B{21}\B{11}, \B{11}\B{20}} \\
        \hline
        3 & {\B{02}\B{12}\B{21}, \B{02}\B{21}\B{11}, \B{02}\B{12}\B{20}} \\ & {\B{12}\B{21}\B{11}, \B{21}\B{11}\B{20}} \\
        \hline
        4 & {\B{02}\B{12}\B{21}\B{11}, \B{02}\B{21}\B{11}\B{20}} \\
        \hline
        5 & {\B{02}\B{12}\B{21}\B{11}\B{20}} \\
        \hline
      \end{tabular}
       }
    \end{minipage}
    \caption{A sample CFG and its blended $n$-gram features}
    \label{fig:node_cfg}
\end{figure}

We noted that the majority (97\%) of indegree and outdegree values were
between 0 and 3, and mainly focused on in-degree values ranging from 0 to 3,
and out-degree values ranging from 0 to 3 when abstracting CFGs.
This abstraction approach results in a total of 16 different basic block types
as shown in the left table of Figure~\ref{fig:node_cfg} where each
entry in the table denotes a specific basic block type annotated with its
indegree and outdegree values.
Basic blocks whose indegree is larger than or equal to 3 are considered as the
same type, and basic blocks whose outdegree is larger than or equal to 3 are
considered as the same type.
During our experiments, we noticed that this approach did not cause significant
feature collision as only 1.27\% of basic blocks from real-world applications
had larger than 3 outdegree and only 3.67\% of basic blocks had larger than 3
indegree.


\subsection{Blended $n$-gram Graphical Feature Extraction}
\label{subsec:feature}


Similar to the traditional $n$-gram analysis, we consider a node (\ie basic block)
in a CFG as a single {\it item}, and define an {\it $n$-gram graphical feature}
to be the {\it consecutive $n$ basic blocks} from an input CFG.
In order to encapsulate the structural properties, the connectivity among nodes,
and the contextual information of the input CFG,
we include all $k$-gram ($k \in [1, n]$) features as the complete graphical
feature set.
This $k$-gram model considering all possible sequences from length $1$ to $n$
was previously referred to as {\it blended} $n$-gram
features~\cite{shawe2004kernel,rieck2008linear}.

Let us take the sample CFG as shown in the middle of Figure~\ref{fig:node_cfg}
to explain the blended $n$-gram graphical features in more details.
Each basic block is represented in the abstracted basic block type as
discussed in Section~\ref{subsec:type}, e.g., \B{21} with 2 parent nodes
and 1 child node.
For a given $n$, we extract all possible blended $n$-gram graphical features at
every node in the CFG.
For example, at node \B{02}, the 1-gram feature is \B{02} itself, the
2-gram features are \B{02}\B{12} and \B{02}\B{21}, and the 3-gram features are
\B{02}\B{12}\B{21}, \B{02}\B{12}\B{20}, and \B{02}\B{21}\B{11}.
This procedure is called {\it visiting} node \B{02}, and visiting a node reaches
descendant nodes at up to $n-1$ levels away.
We apply this procedure for all nodes in the CFG and obtain the resulting
blended $n$-gram graphical feature sets.
The complete 5-gram graphical features for the sample CFG are presented in the
right table of Figure~\ref{fig:node_cfg}.
Note that cycles in a CFG will not be an issue since each node in a CFG is
visited only once and the visiting order makes no difference.


Larger $n$ can result in a larger feature space and provide more distinguishing
capabilities.
On the other hand, a larger feature space also requires more storage and
computing resources to extract all $n$-gram graphical features and perform
subsequent operations, e.g., comparisons.
We comprehensively assessed the impact of different $n$-gram sizes, and
empirically chose blended 5-gram as the default $n$-gram size balancing the
accuracy and the efficiency.
%
A na\"ive implementation of the blended 5-gram feature set would result in a
feature space of 1,118,480\footnote{$16$ of 1-gram features, $16^2$ of
  2-gram features, $16^3$ of 3-gram features, $16^4$ of 4-gram
  features, and $16^5$ of 5-gram features.}.
However, there are certain $n$-gram graphical features
that are invalid by definition.  For example, $k$-gram
($k \geq 2$) features that contain 0 indegree of basic block types
(\ie \B{00}, \B{01}, \B{02}, \B{03}) but do not start with
them are invalid. Similarly, $k$-gram ($k \geq 2$) features that contain
0 outdegree of basic block types (\ie \B{00}, \B{10}, \B{20}, \B{30}) but do not
end with them are also invalid. After removing such invalid
features, the blended 5-gram feature set has smaller 118,096 legitimate entries.

\subsection{Graph Signature Generation and Comparison}
\label{subsec:signature}

We generate a {\it graph signature} by encoding the blended $n$-gram graphical
features into a numeric vector.
Each entry in the vector represents a specific feature, and the value of the
entry denotes the number of appearances of its corresponding feature in the
graph. In this way, both the content and the frequency of features are taken
into consideration when building graph signatures.
All the graph signatures are in the same size which is determined by the
$n$-gram graphical feature space.
We describe a feature entry as a 32-bit unsigned integer type, which we
empirically validate that $2^{32}$ feature space is large enough for practical
usage.

We employ the following cosine similarity measure ($\mathbb{C}$) to compute the similarity
between two graph signatures.

\begin{equation}\label{eq:cos_func}
  \mathbb{C}(\mathcal{G}_a, \mathcal{G}_b) = {{\mathcal{G}_a \cdot \mathcal{G}_b} \over {|\mathcal{G}_a| \cdot |\mathcal{G}_b|}}
\end{equation}

\noindent 
The rationale behind the use of the cosine similarity measure is that
it provides an ideal foundation for effective signature size compression, which
is often desired for large scale analysis. We describe how we generate a more
compact fuzzy hash signature from it in
Section~\ref{sec:fuzzyhash}.


In practice, the feature counts for different binary programs vary
significantly, and the cosine similarity may yield less accurate results as it
only assesses orientation of vectors rather than magnitude of vectors.
For example, two vectors (1,2,3,4) and (2,4,6,8) yield the cosine similarity
score of 1.0, however the magnitude of vectors are significantly different.
This is due to the cosine similarity only measuring the angle between the input
vectors.
We denote the total number of the graphical features contained in a CFG as
$\mathbb{N}$, and define the size rectification factor ($\mathbb{R}$) between
two CFGs as follows:

\begin{equation}\label{eq:rect_length}
  \mathbb{R}(\mathcal{G}_a, \mathcal{G}_b) = {{\mathtt{min}(\mathbb{N}(\mathcal{G}_a), \mathbb{N}(\mathcal{G}_b))} \over {\mathtt{max}(\mathbb{N}(\mathcal{G}_a), \mathbb{N}(\mathcal{G}_b))}}
\end{equation}


We then compute the final similarity ($\mathbb{S}$) between two graph signatures
by multiplying the rectification factor to the cosine similarity,
$\mathbb{S}(\mathcal{G}_a, \mathcal{G}_b) = \mathbb{R}(\mathcal{G}_a,
\mathcal{G}_b) \times \mathbb{C}(\mathcal{G}_a, \mathcal{G}_b)$.
According to the definition, the final similarity of two graph signatures
is 1.0 when they are exactly the same, and the size rectification factor is to
regulate the similarity only when input graph vectors are significantly
different.


\subsection{Fuzzy Hash Signature Generation and Comparison}
\label{sec:fuzzyhash}

The above graph signature representation provides an effective similarity
comparison mechanism; however, the signature size increases exponentially when
larger $n$-gram sizes are used.  For example, the size of the graph signature is
about 461 KB (\ie $4B \cdot 118096$) with blended 5-gram graphical features, which
is challenging to store and compare at a large scale.
To make the technique easier to use and further facilitate the graph signature
storage and comparison process,
we compress the ``raw'' graph signature into $k$-bit vector
representation using fuzzy hashing principles.

Specifically, we pre-define a unique seed number and prepare $k$
independent vectors with elements that are selected randomly from Gaussian
distribution,
and configure each random vector to be the same dimension as
the raw graph signatures.
We denote the {\it i}-th random vector and the raw graph signature as
$\mathcal{V}_i$ and $\mathcal{G}$, respectively; and then define the following
function $\mathbb{B}$ to compare each random vector against the graph
signature.

\begin{equation}\label{eq:sgn_func}
  \mathbb{B}_i(\mathcal{V}_i, \mathcal{G}) = \begin{cases}
1 \quad \text{if $\mathcal{V}_i \cdot \mathcal{G} \geq 0$} \\
0 \quad \text{if $\mathcal{V}_i \cdot \mathcal{G} < 0$ } \\
\end{cases}
\end{equation}

In this way, each random vector is used to create one projection for
the raw graph signature based on the dot product between
the random vector and the graph signature, and the output
of $k$ times of projections becomes a $k$-bit vector.
The overall random projection procedure is formally known as
hyperplane locality sensitive hashing (LSH)~\cite{charikar2002similarity,datar2004locality}.
Since all the graph signatures are projected into \{0, 1\} space
through the same hashing process, graph signatures with close ``locality'' will
be projected to similar $k$-bit vectors.
We consider the generated $k$-bit hash value as a {\it fuzzy hash signature}.
Given two fuzzy hash signatures $\mathcal{F}_a$ and $\mathcal{F}_b$,
we compute Hamming similarity ($\mathbb{H}$) to measure the similarity
between the projected hash outputs as follows:

\begin{equation}
  \mathbb{H}(\mathcal{F}_a, \mathcal{F}_b) = 1 - {|{\mathcal{F}_a} \oplus {\mathcal{F}_b}| \over k}
\end{equation}




LSH is commonly used as an efficient technique to conduct the approximate nearest
neighbor search in high-dimension objects.
Previous research in LSH
domain~\cite{gionis2001efficient,charikar2002similarity} showed that
(1) cosine similarity (as used for graph signature comparison) is one type of similarity
measurement that admits LSH families; and
(2) for any similarity function $sim(x, y)$ that admits LSH projection,
we can always obtain an LSH family that maps the original objects to \{0, 1\}
space and has the property that the similarity between projected objects (\eg
Hamming similarity) is proved to correspond to the original similarity function
at ${{1+sim(x, y)} \over 2}$.
Therefore, we leverage $\mathbb{H}(\mathcal{F}_a, \mathcal{F}_b)$ to estimate
the original cosine similarity between graph signatures by:

\begin{equation}\label{eq:sim_hash}
  \tilde{\mathbb{C}}(\mathcal{G}_a, \mathcal{G}_b) = {2 \times
    \mathbb{H}(\mathcal{F}_a, \mathcal{F}_b) - 1}
\end{equation}


Increasing the number of random projection vectors makes the similarity estimation more accurate.
To obtain the optimal n-gram size and fingerprint size, we conduct grid search against the
same ground truth dataset with a set of potential parameters, and record
the optimal clustering result considering both the efficiency and accuracy\footnote{
the parameter selection process is not included in the paper due to space limitation}.
According to the empirical process, we set 256 as the optimal $k$ value.
The total number of graphical features $\mathbb{N}$ is
represented as a 32-bit integer, and the final fuzzy hash output is 288 bits by
default.
We compute the final fuzzy hash similarity by multiplying
the rectification factor to the estimated hash similarity:
$\tilde{\mathbb{S}}(\mathcal{G}_a, \mathcal{G}_b) = \mathbb{R}(\mathcal{G}_a,
\mathcal{G}_b) \times \tilde{\mathbb{C}}(\mathcal{G}_a, \mathcal{G}_b)$.



\section{Evaluation of CFG Similarity Analysis Algorithms}
\label{sec:experimentation}


In this section, we evaluate the effectiveness and accuracy of different CFG
comparison algorithms.
Our evaluation mainly focuses on the capability of the algorithms
to differentiate CFG structures (\ie the topology) {\it without} relying on
basic block content.
The auxiliary information provided by basic block content
could further improve CFG comparison. For example, the abstraction process
discussed in Section~\ref{subsec:type} can be extended to incorporate such
information.

We compared \fuzzname to representative CFG similarity analysis algorithms
discussed in Section~\ref{sec:related}.
For min-cost bipartite graph matching
algorithm~\cite{hu2009large,vujovsevic2013software},
{\it k}-subgraph matching~\cite{kruegel2005polymorphic}, and simulation-based
graph comparison~\cite{sokolsky2006simulation}, we used the implementations
provided by Chan~\cite{chan2014method}.
We implemented McGreger's maximum common subgraph matching algorithm~\cite{mcgregor1982backtrack} and our \fuzzname.
The graph embedding based CFG comparison algorithms~\cite{feng2016scalable,xu2017neural}
are not evaluated since they rely
upon a separate training process and require six types of specific features
derived from concrete basic block content.
The final outputs of the algorithms were normalized ranging from 0 to 1.
To facilitate evaluating arbitrary CFG comparison algorithms,
we plan to release our evaluation framework and the corresponding dataset.
A new CFG comparison algorithm can be easily evaluated in this framework
by providing a plugin that takes two CFGs as input and outputs a normalized similarity score.

\subsection{Algorithm Evaluation Strategy}
\label{subsec:eval}

To the best of our knowledge, the only prior work that formally
evaluated different CFG similarity algorithms was that of Chan
\etal~\cite{chan2014method}.
They created a ground truth CFG dataset by applying different levels of edit operations to
a seed CFG, and checked if the algorithms could identify a similar level of
similarity differences between the generated testing CFGs and the seed CFG.
However, we observe that Chan \etal's methodology is problematic from the following perspectives:
(1) the ground truth dataset and evaluating strategy
were inherently biased towards edit-distance based CFG comparison algorithms;
(2) different edit operations (\eg adding/deleting a node, and adding/deleting
an edge)
might have different costs. For example, editing a node will not impact any
existing edges; on the contrary, editing an edge will affect two nodes.
Therefore, the testing CFGs generated by the same number of edit operations may
present different similarity levels.



We propose a new evaluation strategy
where we employ a CFG comparison algorithm in a {\it hierarchical agglomerative
clustering (HAC) system} as a custom distance function,
then use the custom distance function to conduct clustering analysis
for the same ground-truth dataset, and use the overall clustering result
as a performance indicator of the corresponding CFG comparison algorithm.
HAC is a bottom-up version of the hierarchical clustering methods,
in which all input items are initially considered as singleton
clusters, and then for a specified distance threshold {\it t}
the algorithm iteratively merges the clusters with the minimum distance
as long as the corresponding cluster distance {\it d} is less than {\it t}.
The distance between two clusters is often referred to as ``linkage''
and the following three linkage criteria are
commonly used: {\it single linkage} considering the cluster
distance as the minimum distance between all the entries of two clusters,
{\it average linkage} considering the cluster distance
as the average distance between all the entries of two clusters, and
{\it complete linkage} considering the cluster distance as
the maximum distance between all the entries of two clusters.

The rationale for evaluating different CFG comparison algorithms
through HAC are that:
(1) the fundamental component of a HAC system is the similarity measurement
between all input items, which can be pre-calculated as a distance matrix using
each CFG comparison algorithm;
(2) when analyzing the same ground truth dataset,
the only parameter that will impact the final clustering result
is the distance matrix which is controlled by each CFG comparison algorithm;
(3) the clustering analysis procedure assesses the capability to
group similar items and separate different items at the same time.


To measure the clustering results, we adopt the measurement of precision and recall.
%
%
Precision and recall measure two competing criteria of a clustering
algorithm: the ability to separate items from different clusters, and the ability
to group together items belonging to the same cluster.
We consider the intersection point (or the nearest point) between precision and recall
to be the optimal clustering output.
For simplicity, all the clustering results are subsequently measured with F-score,
which is the harmonic mean of the optimal precision and recall.

\subsection{Experiment Data Preparation}

To create a ground-truth CFG dataset, we compiled the latest version of the
Android Open Source Project code and obtained 588 ELF ARM64 binaries.
We analyzed the compiled binaries, collected all the function level CFGs
that had 20 nodes, and then randomly selected 5 seed CFGs from all
available 20-node CFGs. We applied {\it one} edit operation (\eg adding a node,
deleting a node\footnote{a graph node can be deleted only if it is isolated},
adding an edge, and deleting an edge) for each seed CFG.
Since all the edges and nodes can be edited in single operation and any of the
existing two nodes can be added with an additional edge, each of the seed CFG
can be used to create about 500 artificial and structurally similar CFGs.
In the end, we created a total of 1,934 CFGs from the 5 seed CFGs. 
We selected CFGs with 20 nodes since they typically provided enough varieties
between different CFGs, and the individual CFG comparison did not take
too long to complete for all the evaluated algorithms.


\subsection{Evaluation Results}

\begin{figure*}[t]
  \centering
  \includegraphics[width=\linewidth,height=.25\textheight]{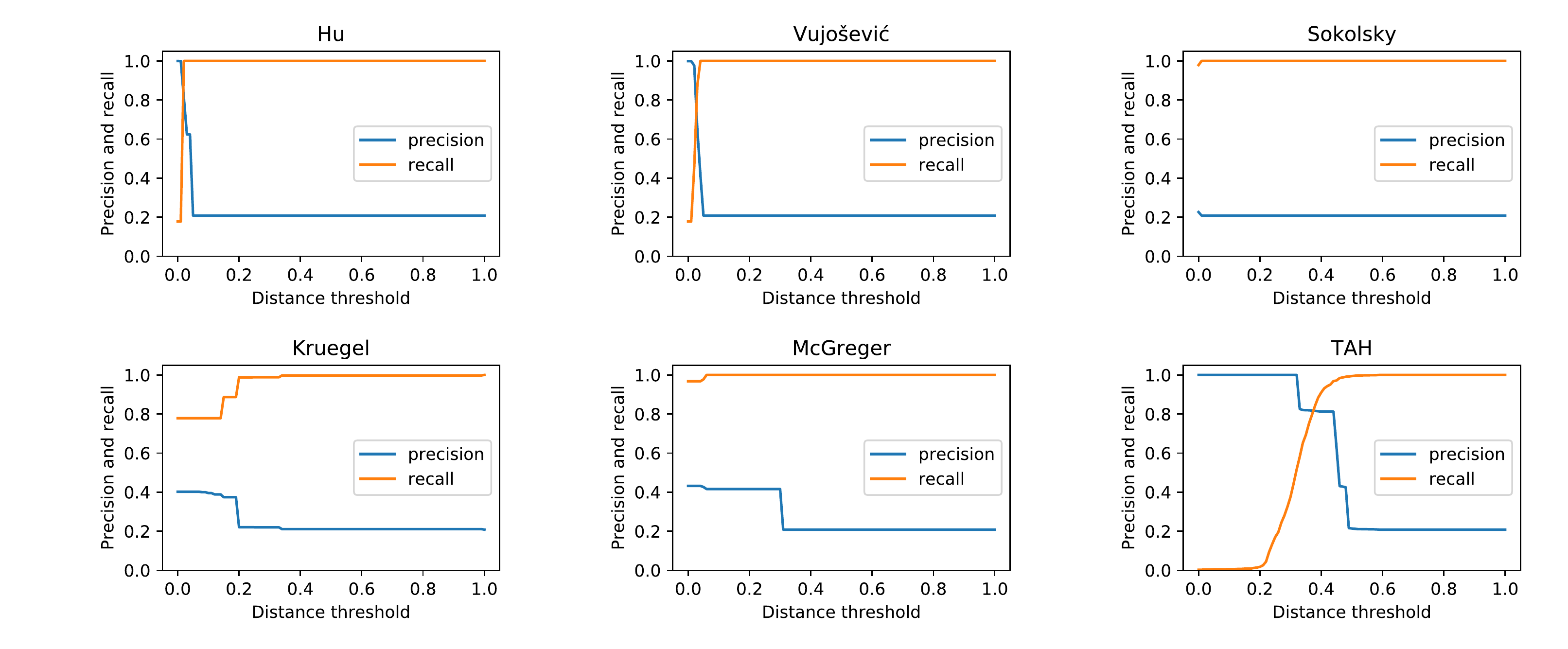}
  \centering
  \caption{Single-linkage clustering results for different CFG algorithms}
  \label{fig:sim_res1}
\end{figure*}

To avoid the bias towards a particular linkage strategy, we report the
clustering results with three linkage approaches for all CFG comparison algorithms.
We summarize the optimal clustering results for different algorithms in
Table~\ref{table:cfg_clustering}, and present the detailed single-linkage
CFG clustering results in Figure~\ref{fig:sim_res1}.
The fuzzy hash signature based CFG comparison approach is labeled as \fuzzname
and the graph signature based CFG comparison approach is labeled as \algoname.
We included \algoname to verify that fuzzy hashing based \fuzzname closely
approximated \algoname with little impact on accuracy.

\begin{table*}[hh]
\caption{Optimal clustering results for different CFG comparison algorithms}
\label{table:cfg_clustering}
\centering
\begin{tabular}{|c||c|c|c|c|}
  \hline
{Algorithm}  & {Single linkage} &  {Average linkage} &  {Complete linkage} & {Avg F-score} \\
\hline
\hline
  Hu~\cite{hu2009large} & 0.847 & 0.872 & 0.879 & 0.866 \\
  \hline
  Vujošević~\cite{vujovsevic2013software} & 0.749 & 0.869 & 0.876 & 0.831 \\
  \hline
  Sokolsky~\cite{sokolsky2006simulation} & 0.367 & 0.456 & 0.501 & 0.441\\
  \hline
  Kruegel~\cite{kruegel2005polymorphic} & 0.530 & 0.530 & 0.530 & 0.530\\
  \hline
  McGreger~\cite{mcgregor1982backtrack} & 0.597 & 0.588 & 0.324 & 0.503\\
  \hline
  \algoname & 0.816 & 0.926 & 1.000 & 0.914\\
  \hline
\fuzzname & 0.817 & 0.926 & 0.864 & 0.869\\
\hline
\end{tabular}
\end{table*}

\begin{figure}[ht]
    \centering
    \begin{subfigure}{0.5\linewidth}
        \centering
        \includegraphics[width=\linewidth, height=0.17\textheight]{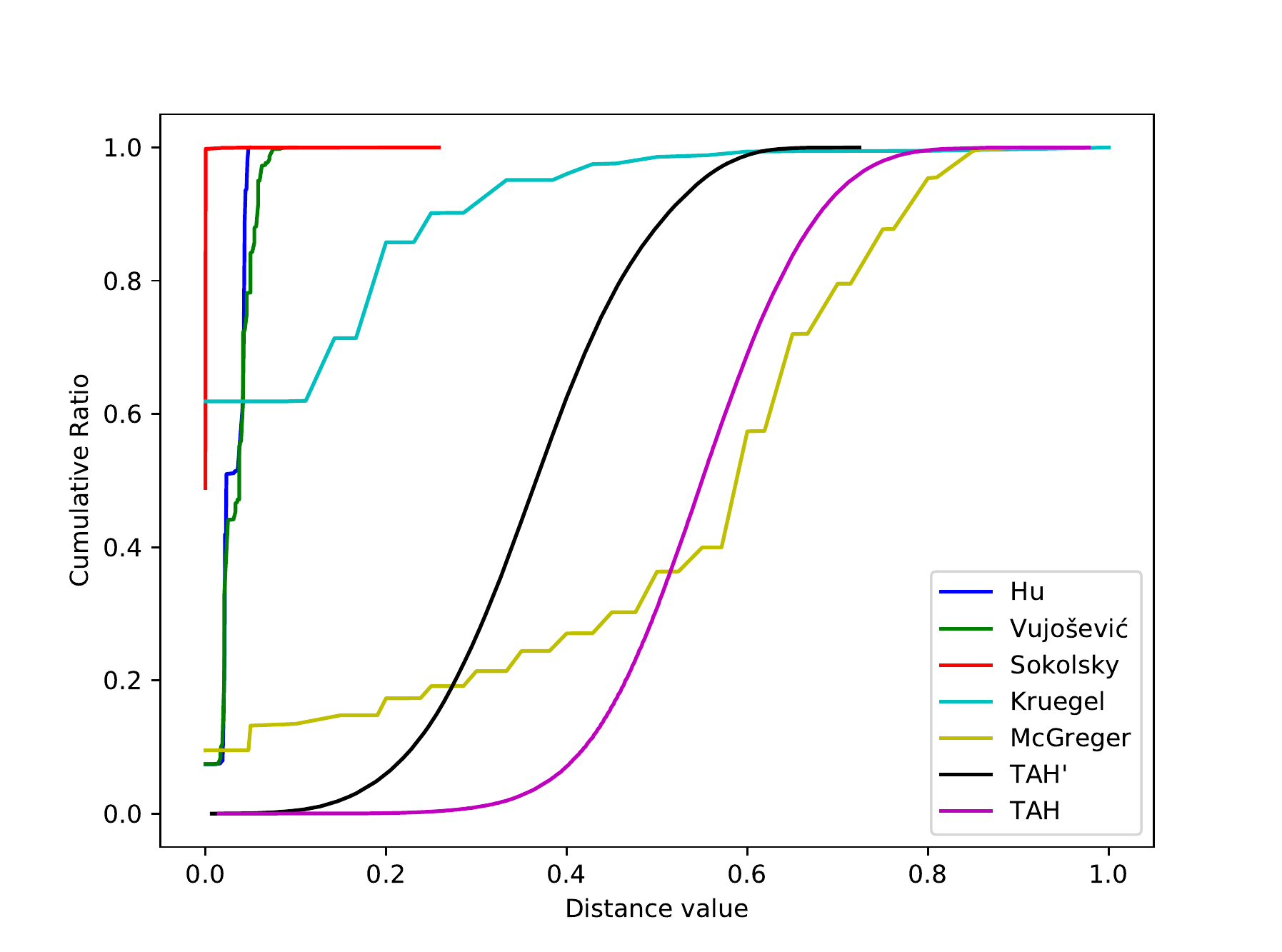}
        \caption{Same-group CFG pairs}
        \label{fig:same}
    \end{subfigure}%
    \begin{subfigure}{0.5\linewidth}
        \centering
        \includegraphics[width=\linewidth, height=0.17\textheight]{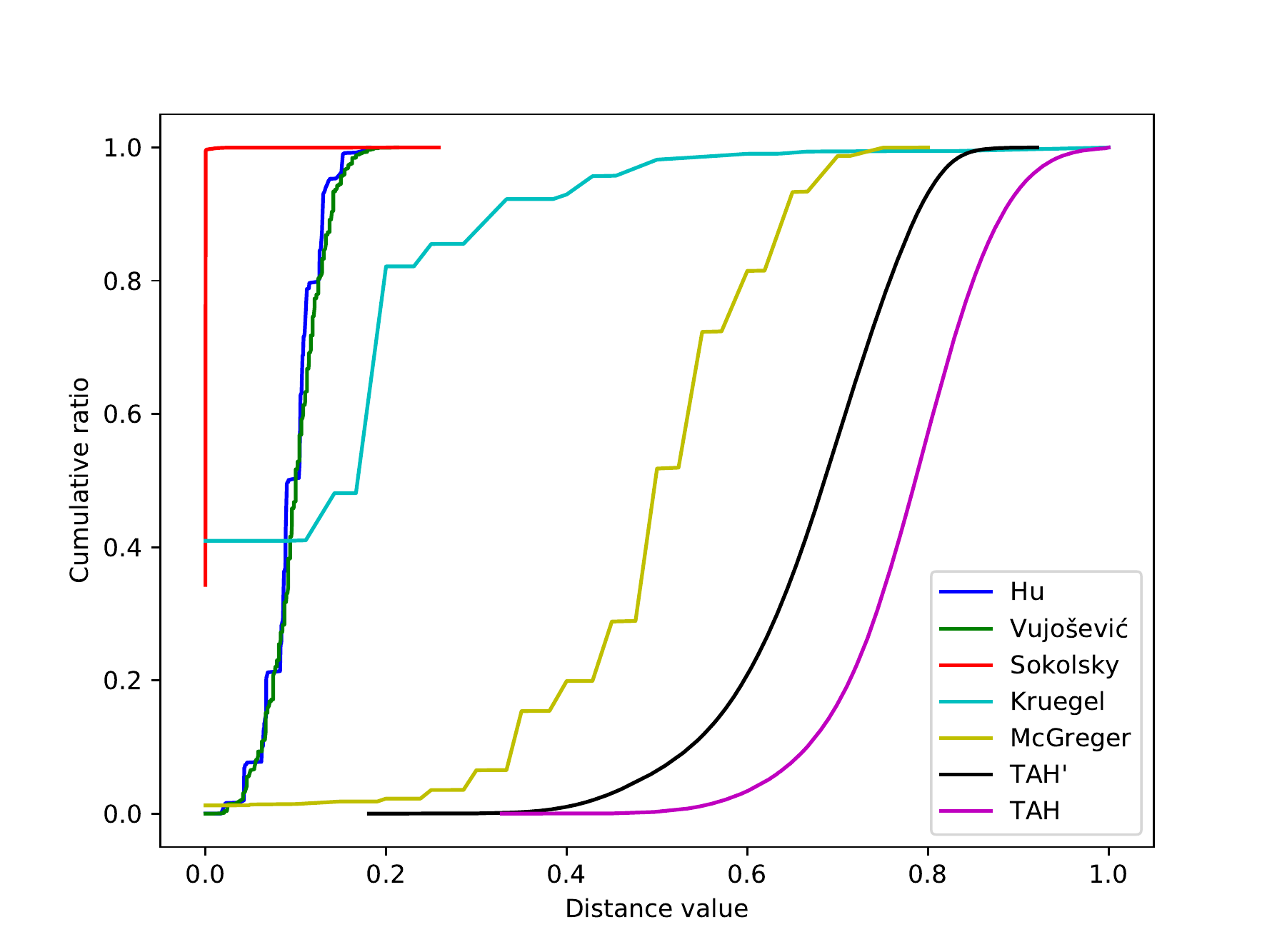}
        \caption{Different-group CFG pairs}
        \label{fig:diff}
    \end{subfigure}%
    \caption{Cumulative distance distribution for all comparison algorithms}
    \label{fig:cdf}
\end{figure}

To further dissect the clustering results, we separated all CFG
pairs into two categories: same group CFG pairs and different group CFG pairs.
Ideally, the distance of the same group CFG pair is expected to be small,
and the distance of the different group CFG pair is expected to be large.
We present the minimum and the maximum distances of each group in
Table~\ref{table:distance}, and plot the cumulative distance distribution
for each algorithm in Figure~\ref{fig:cdf}.

\begin{table*}[h]
\caption{Distance ranges for different CFG pairs}
\label{table:distance}
\centering
\scalebox{0.965}{
\begin{tabular}{|c||c|c|c|c|c|c||c|c|c|c|c|c|}
\hline
  \multirow{2}{*}{Algorithm} & \multicolumn{2}{c|}{Same-group} & \multicolumn{2}{c|}{Diff-group} & \multicolumn{2}{c||}{All pairs} & \multicolumn{2}{c|}{Same-group} & \multicolumn{2}{c|}{Diff-group} & \multicolumn{2}{c|}{All groups}\\ \cline{2-13}
& {   Min   } & {   Max   }& {   Min   } & {   Max   }& {   Min   } & {   Max   } & {   Min   } & {   Max   }& {   Min   } & {   Max   }& {   Min   } & {   Max   } \\
\hline
\hline
Hu & 0.000 & 0.049 & 0.000 & 0.182 & 0.000 & 0.182 & 0.000 & 0.032 & 0.029 & 0.165 & 0.000 & 0.165 \\
Vujošević & 0.000 & 0.100 & 0.000 & 0.213 & 0.000 & 0.213 & 0.000 & 0.067 & 0.028 & 0.203 & 0.000 & 0.203 \\
Sokolsky & 0.000 & 0.258 & 0.000 & 0.258 & 0.000 & 0.258 & 0.000 & 0.250 & 0.000 & 0.250 & 0.000 & 0.250 \\
Kruegel & 0.000 & 1.000 & 0.000 & 1.000 & 0.000 & 1.000 & 0.000 & 1.000 & 0.000 & 1.000 & 0.000 & 1.000 \\
McGreger & 0.000 & 0.905 & 0.000 & 0.800 & 0.000 & 0.905 & 0.000 & 0.935 & 0.000 & 0.833 & 0.000 & 0.935 \\
\algoname & 0.007 & 0.724 & 0.181 & 0.921 & 0.007 & 0.921 & 0.000 & 0.777 & 0.294 & 0.917 & 0.000 & 0.917 \\
\fuzzname & 0.015 & 0.978 & 0.328 & 1.000 & 0.015 & 1.000 & 0.000 & 0.964 & 0.398 & 1.000 & 0.000 & 1.000 \\
\hline
\end{tabular}
}
\end{table*}

Combining the distance range information with CFG clustering results, we can see that:
(1) the algorithms proposed by Hu, Vujošević, and Sokolsky had very narrow
overall distance ranges for all CFG pairs, thus CFGs were quickly merged
into one group during the clustering process, which resulted in high recall
and low precision for the majority of provided distance thresholds.
However, the edit distance based CFG comparison algorithms
generated relatively lower distance outputs for same group CFG pairs and higher
distance outputs for different group CFG pairs, therefore they still generated
overall good F-score outputs.
(2) the algorithms proposed by Kruegel and
McGreger provided a broader distance range,
but for both same group CFG pairs and different group CFG pairs at the same
time.
This made precision and recall from different clustering thresholds
slowly intersected with each other or never intersected at all, which resulted
in overall poor F-score.
(3) \algoname and \fuzzname both provided very good distance ranges.
Figure~\ref{fig:cdf} also demonstrates that they clearly separated
majority of same group CFG pairs and different group CFG pairs,
\eg when choosing the distance threshold around 0.50 for \algoname and choosing
the distance threshold around 0.70 for \fuzzname.
Therefore, they both generated very good F-score outputs.

As mentioned earlier, all the algorithm implementations evaluated in this paper
only considered the topology of the CFGs and ignored the content of basic
blocks,
\ie the content similarity between all basic block pairs were considered as 1.
Therefore, these implementations may not faithfully represent the full
capability of the original designs, and the evaluation results presented in this
paper only reflected
the algorithms' capability of measuring the similarity of CFG structures,
without considering the basic block content.
Even though our evaluation framework supports experiments with real-world CFGs
and the \fuzzname can be extended to incorporate boolean node
attributes\footnote{\eg adding 1 bit to record whether the node contains string
  constants during basic block type abstraction as described in
  Section~\ref{subsec:type}}, it is practically challenging to conduct large
scale experiments with real-world CFGs as it is nontrivial to prepare
a large-scale ground truth dataset using real-world CFGs that
present {\it controlled} and {\it known} similarity levels.
Firstly, the source code level similarities are commonly not proportionally preserved
after the complicated compilation procedure, \ie it is difficult to control
the granularity of CFG similarities through source code updates.
Secondly, when analyzing large amount malicious real-world binary samples that
were labeled as the same malware family label, we noticed that samples often
either shared the exact same functions or had significantly changed functions,
while few CFGs were closely similar.

Our evaluation strategy highlighted the strengths and the weaknesses of
existing CFG comparison algorithms when comparing graph topologies.
In summary, \algoname showed the
best separation capabilities between similar CFG pairs and different CFG pairs,
and \fuzzname was a faster approximate method
yet still quite comparable to the accuracy of \algoname,
whereas existing CFG comparison algorithms either had limited distance output
ranges or had almost the same distance ranges
between same group CFG pairs and different group CFG pairs.
This demonstrates that \algoname and \fuzzname are more reliable and have more
balanced capability to recognize similar CFGs and identify different CFGs at the
same time.
High F-scores of \algoname and \fuzzname are achieved mainly because
the differences between $n$-gram graphical features
are always proportional to the structural differences of CFGs.
For example, the same group CFGs will have more similar $n$-gram graphical
features, and the different group CFGs will have more different graphical
features, which subsequently leads to the desired distinguishing capabilities.

\subsection{Overall Performance}


For each algorithm, we measured the time taken (in seconds) to finish the similarity
calculation for all CFG pairs.
Since the hierarchical clustering algorithm has $n^2$
complexity, the prepared 1,934 CFGs would result in 1,869,211 pairwise CFG comparisons.
Note that all the algorithm implementations took two CFGs as input
and produced a similarity score, which means the graph signatures
for \algoname and fuzzy hash signatures for \fuzzname were
generated 1,869,211 times during the evaluation.
And since \fuzzname fuzzy hash outputs are generated from \algoname graph signatures,
this made the \fuzzname approach dramatically slower than
other approaches.
However, if we cache the graph embedding process, \eg pre-calculating
the graph signature for \algoname and the fuzzy hash output for \fuzzname for each CFG
then directly loading the generated signatures/hashes within each pairwise CFG comparison routine,
\algoname only took 125.5s to
finish the graph signature generation and all pairwise comparison,
and \fuzzname only took 23.8s to finish the fuzzy hash generation and
all pairwise comparison.
This was much more efficient than other existing approaches.
For the same dataset, Hu's algorithm took 755.6s, Vujošević's algorithm took 1788.0s,
Sokolsky's algorithm took 483.1s, Kruegel's algorithm took 321.8s, and
McGreger's algorithm took 2542.1s to finish.
Note that for the same CFG input, the fuzzy hash output is the same if we apply the algorithm
and generate it again, so it makes sense to only generate it once and
caching the intermediate and compact CFG representation when used in real-world.
However, this won't be applicable for the existing algorithms since there is
no intermediate CFG representation for them and they need to repeated compare the ``raw'' CFG inputs.
This evaluation indicates that \fuzzname is particularly suitable for large scale
dataset analysis.





\section{Evaluation of Binary Similarity Analysis Tools}
\label{sec:application}

Based on \fuzzname, we implemented a binary similarity analysis
tool\footnote{For simplicity, we also refer our binary similarity analysis tool
  as \fuzzname.} to evaluate
the effectiveness of the structural comparison approach.
We assessed the effectiveness of \fuzzname
to conduct binary similarity analysis, and compared it with the following
existing binary similarity comparison solutions:
SSdeep v2.14.1~\cite{kornblum2006identifying} and BinDiff v4.3.0~\cite{dullien2005graph}.
Other previously proposed binary similarity analysis tools were not evaluated because they
were neither maintained~\cite{bourquin2013binslayer}
(\ie not working with a majority of the collected binaries) nor publicly
available~\cite{gao2008binhunt}.
In order to exclude the potential impact of the different disassemblers on the
CFG construction accuracy, \fuzzname was implemented as an IDA Pro plugin.
The current implementation of \fuzzname used IDA Pro v6.8
to process the target binaries and constructed the corresponding CFGs.
The same version of IDA Pro was also used by BinDiff.

We embedded all the binary similarity analysis tools (\eg \fuzzname, BinDiff,
and SSdeep)
into the hierarchical clustering system, and used the clustering outputs to
evaluate the similarity measurement accuracy of the tools.
We prepared the ground truth dataset by collecting desktop
malware samples labeled with malware family names, and considered that the
binaries with the same family name were more similar to each other than binaries
from different families were.
Because the labeled malware datasets used in previous research
were either discontinued or only contained a list of file hashes, we prepared
our own labeled ground truth malware dataset.
In the end, we collected 2,865 recent desktop malware samples from
VirusShare~\cite{virusshare}, and every sample was consistently labeled by
at least 25 antivirus products listed on VirusTotal~\cite{virustotal}.
The resulting ground truth malware dataset containing 8 different malware
families is shown in Table~\ref{table:mal_data}.

\begin{table}[h]
\caption{Ground truth malware dataset}
\label{table:mal_data}
\centering
\begin{tabular}{|c|c||c|c||c|c|}
\hline
Malware Family & Size & Malware Family & Size  & Malware Family & Size \\
\hline
\hline
InstalleRex & 1115 & OutBrowse  & 615 & MultiPlug & 384 \\
\hline
DomaIQ & 184 & LoadMoney & 173 & Linkular & 164 \\
\hline
InstallCore & 127 & DownloadAdmin & 103 & & \\
\hline
\end{tabular}
\end{table}
\vspace{-.1 in}

\begin{figure}[t]
  \includegraphics[width=\linewidth,height=0.37\textheight]{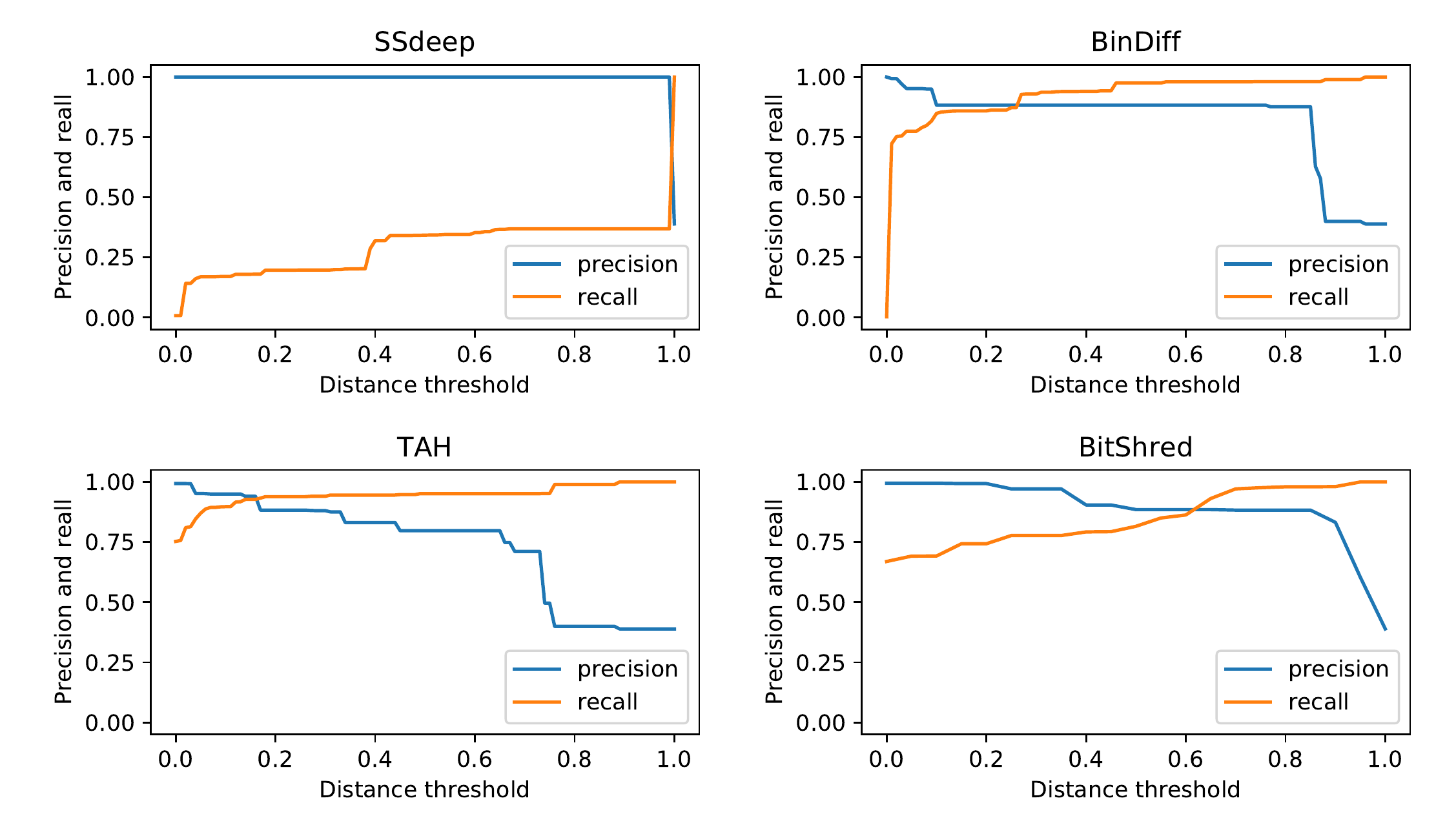}
  \centering
  \caption{Single-linkage clustering results of different similarity analysis tools}
  \label{fig:bin_acc}
\end{figure}

We also measured precision and recall to evaluate the clustering outputs.
For all tools, we summarized the optimal clustering
results with regard to different clustering strategies in Table~\ref{table:bin_acc},
and depicted the overall clustering results with different binary similarity
analysis tools in Figure~\ref{fig:bin_acc}.
We can see that \fuzzname generated the
highest F-score of 0.929 for single linkage clustering analysis.
BinDiff produced similar results with F-score of 0.883,
while SSdeep only achieved overall F-score of 0.690.
Because SSdeep operated at the binary stream level, it was not able to identify
a significant number of semantically similar binaries. BinDiff and \fuzzname
both operated at the CFG level, and effectively identified a larger number of
similar binaries.
Different similarity calculation logic of the tools led to the F-score
differences between BinDiff and \fuzzname.

\vspace{-.1 in}

\begin{table}[h]
\caption{Optimal clustering results for different binary similarity analysis tools}
\label{table:bin_acc}
\centering
\begin{tabular}{|c||c|c|c||c|}
\hline
Tool & {Single linkage}& {Average linkage}&  {Complete linkage} & {Time taken} \\
\hline
\hline
{   SSdeep   } & 0.690 & 0.690 & 0.689 & 2.7m \\
{    BinDiff    } & 0.883 & 0.883 & 0.883 & 166.4m \\
\fuzzname & 0.929 & 0.903 & 0.909 & 0.9m \\
\hline
\end{tabular}
\end{table}

\vspace{-.1 in}

We also evaluated the time taken to conduct the experiments with each tool.
Since BinDiff did not have an intermediate ``signature'' representation for
binary CFGs,
we only calculated the time it took to finish all the pairwise computations by
converting all the input binaries into {\tt BinExport} format.
The time to conduct pairwise comparisons with each tool is shown in the last
column of Table~\ref{table:bin_acc}.
We can see that \fuzzname outperformed SSdeep and BinDiff in terms of
efficiency, and BinDiff was dramatically slower than other tools.
The main reason for the higher efficiency of \fuzzname was that
a fuzzy hash signature was essentially a bit-vector which was more CPU
friendly than SSdeep hash representation was.

It is worth to mention that various previous malware clustering systems already
demonstrated very promising results.
For example, using different datasets, Malheur~\cite{rieck2011automatic} reported F-score of 0.95,
BitShred~\cite{jang2011bitshred} showed F-score of 0.932, and
FIRMA~\cite{rafique2013firma} claimed F-score of 0.988.
As a reference, we chose to conduct experiments with BitShred, which
was a state-of-the-art malware clustering tool using static and dynamic binary
features.
Since BitShred only adopted the single-linkage clustering strategy, we plot
the single-linkage clustering results for all the tools in
Figure~\ref{fig:bin_acc}.
From the graph, we can see that BitShred reached the optimal F-score of 0.885.
Overall, \fuzzname generated the best clustering results (\eg F-score of 0.929)
with single-linkage clustering.
Figure~\ref{fig:bin_acc} also shows that recall for \fuzzname and BitShred
at the distance threshold of 0 are above 0.650,
while recall of SSdeep and BinDiff at this threshold was 0.
This is because \fuzzname and BitShred correctly identified a significant number
of binary pairs as similarity of 1.000, while both SSdeep and BinDiff
could not identify any of such binary pairs.
We further notice that the majority of precision values (\eg with
distance thresholds of [0.000, 0.995]) for SSdeep
were 1.000, which means all the binary pairs that were identified as
similar (\ie similarity score larger than 0.005) were indeed similar.
However, at the same time, the corresponding recall values for SSdeep were less
than 0.365,
which indicates that SSdeep failed to recognize a significant number of binaries
that were known to be similar regardless of the distance thresholds.
This is in line with our practical usage experience with SSdeep.

\section{Limitation}
\label{sec:limitation}






\subsection{Feature and Signature Collision}

We consider the potential attack scenario by generating
similar graph features for different CFGs.
According to the design in Section~\ref{subsec:type}, feature
collision happens when: (1) nodes have indegree or outdegree larger than 3;
(2) nodes with different content are in the same topology.
The first type of collision is rare in real-world binary programs (\eg
3.67\% of nodes in our real-world datasets), and
the second type of collision is largely alleviated by recording the
node context (\ie $n$-gram features) and graphical feature counts.
$n$-gram feature extraction allows the resulting feature
differences to be proportional to the structural differences.
The proposed CFG similarity analysis algorithm \fuzzname compares CFG graph
signatures by measuring the overall cosine similarity
of graph signatures and the relative CFG sizes.
Since the graph signature is mainly a summary of all features contained in a
CFG, it is theoretically possible for binaries with different CFGs
to generate similar graph signatures.
However, random signature collision for overall binary CFGs is rare in practice.
To further reduce the possibility of collision, we can
increase the graphical feature space by incorporating certain
basic block content information into the type abstraction process,
such as the presence of a string or numeric constant, and the number of
instructions.


\subsection{Obfuscation and Evasion Techniques}

It is well-known that malware samples are often packed in recent years to evade signature-based
malware analysis tools~\cite{guo2008study}.
Even worse, malware authors can apply multiple layers of packing,
or employ advanced packers that dynamically decrypt original code on-the-fly
or interpret instructions in a virtualized environment.
%
%
We believe the \fuzzname-based binary similarity analysis
is still useful in practice because of the following reasons.
(1) Lots of real-world binaries are still unpacked, especially for adware or PUP
programs.
\fuzzname can be used to quickly filter out similar binaries that
have been processed before, or used for triaging a large number of unprocessed
binaries (even packed ones) by grouping similar instances together.
For this purpose, traditional cryptographic hash and existing similarity
analysis solutions are less effective.
(2) CFG level analysis makes it possible to provide a binary similarity analysis
solution that has a good balance between accuracy and efficiency.
For example, dynamic analysis based approach can defeat obfuscation, but is
almost impossible to efficiently analyze a large scale of dataset.
(3) Comparing to low-level binary sequences, it will be more difficult
to add randomness to the CFG structure for the packed binaries.
For instance, the dead code can be removed during the CFG construction procedure.
Thus the packed binaries would often share certain deobfuscation
routines, which can be viewed as the signature of the packers.



\section{Conclusion}
\label{sec:conclusion}

In this paper, we proposed an effective CFG comparison algorithm
\fuzzname, which compares CFGs using $n$-gram graphical features.
In order to compare with existing CFG comparison solutions, we designed a
clustering analysis based evaluation framework, and
systematically showed that \fuzzname was more accurate and efficient
compared to state-of-the-art CFG comparison techniques.
Based on \fuzzname, we also developed a graphical comparison based fuzzy
hash tool for binary similarity analysis.
We empirically demonstrated that \fuzzname outperformed existing binary
similarity analysis tools
while conducting malware clustering analysis.



\section{Acknowledgment}
\label{sec:acknowledgment}

This research was partially supported by the U.S. National Science
Foundation under grant No. 1622402 and 1717862. Any opinions, findings
and conclusions or recommendations expressed in this material are
those of the authors and do not necessarily reflect the views of the
National Science Foundation.


{\normalsize \bibliographystyle{acm}
\bibliography{paper}}



\end{document}